\documentstyle[prl,multicol,amsmath,aps,psfig]{revtex}
\begin{document}

\title{ 
Monte Carlo Dynamics of driven Flux Lines in Disordered Media}
\author{Alberto Rosso and Werner Krauth
\footnote{rosso@lps.ens.fr; krauth@lps.ens.fr,
http://www.lps.ens.fr/$\tilde{\;}$krauth} 
}
\address{CNRS-Laboratoire de Physique Statistique \\
Ecole Normale Sup{\'{e}}rieure,
24, rue Lhomond, 75231 Paris Cedex 05, France}
\maketitle
\begin{abstract} 
We show that the common local Monte Carlo rules used to simulate
the motion of driven flux lines in disordered media cannot capture
the interplay between elasticity and disorder which lies at the heart of
these systems. We therefore discuss a class of generalized Monte Carlo
algorithms where an arbitrary number of line elements may move at
the same time. We prove that all these dynamical rules have the
same value of the critical force and possess  phase spaces made up
of a single ergodic component. A variant Monte Carlo algorithm
allows to compute the critical force of a sample in a single pass
through the system. We establish dynamical scaling properties and
obtain precise values for the critical force, which is finite even
for an unbounded distribution of the disorder.  Extensions
to higher dimensions are outlined.

\end{abstract}

\begin{multicols}{2}
\narrowtext

In the last few years, the study of elastic manifolds in random
media has retained much attention.

These systems appear in a wide range of physical systems, ranging
from vortices in type-II superconductors \cite{supra}, to charge
density waves \cite{CDW}, interfaces in disordered magnets
\cite{Lemerle}, and to the problem of directed polymer growth
\cite{KPZ}.  The response of elastic manifolds to an external
driving force $f$ is highly non-trivial: at temperature $T=0$, the
manifold is completely ``pinned'' at small forces, while it moves
with non-zero velocity at forces larger than a certain critical
force $f_c$. At finite, but small,  $T$, a socalled ``creep motion''
takes place for $f \ll f_c$, while the motion at $f \gg f_c $ is
described by viscous flow.  Many details of this dynamical problem,
both at $T=0$ and at finite temperatures, have yet to be understood
fully \cite{Chauve,Kardar}.

This paper is concerned with an analysis of the dynamical Monte
Carlo method \cite{Intro} as applied to  lattice models of driven
elastic manifolds in random media.  We argue that the common local
Monte Carlo rules \cite{Ji,Roters1,Roters2,Yoshino} are incompatible
with the Langevin dynamics \cite{Marchetti,Nattermann1,Nattermann2}
 which defines time evolution in continuum models.  We instead
propose generalized Monte Carlo algorithms where an arbitrary number
of  elements may move at the same time.  For this class of algorithms,
we can establish the uniqueness of the critical force and simple
connectedness of phase space.  Furthermore, we devise a method
which simplifies enormously the calculation of the critical force.

\begin{figure}
\centerline{ \psfig{figure=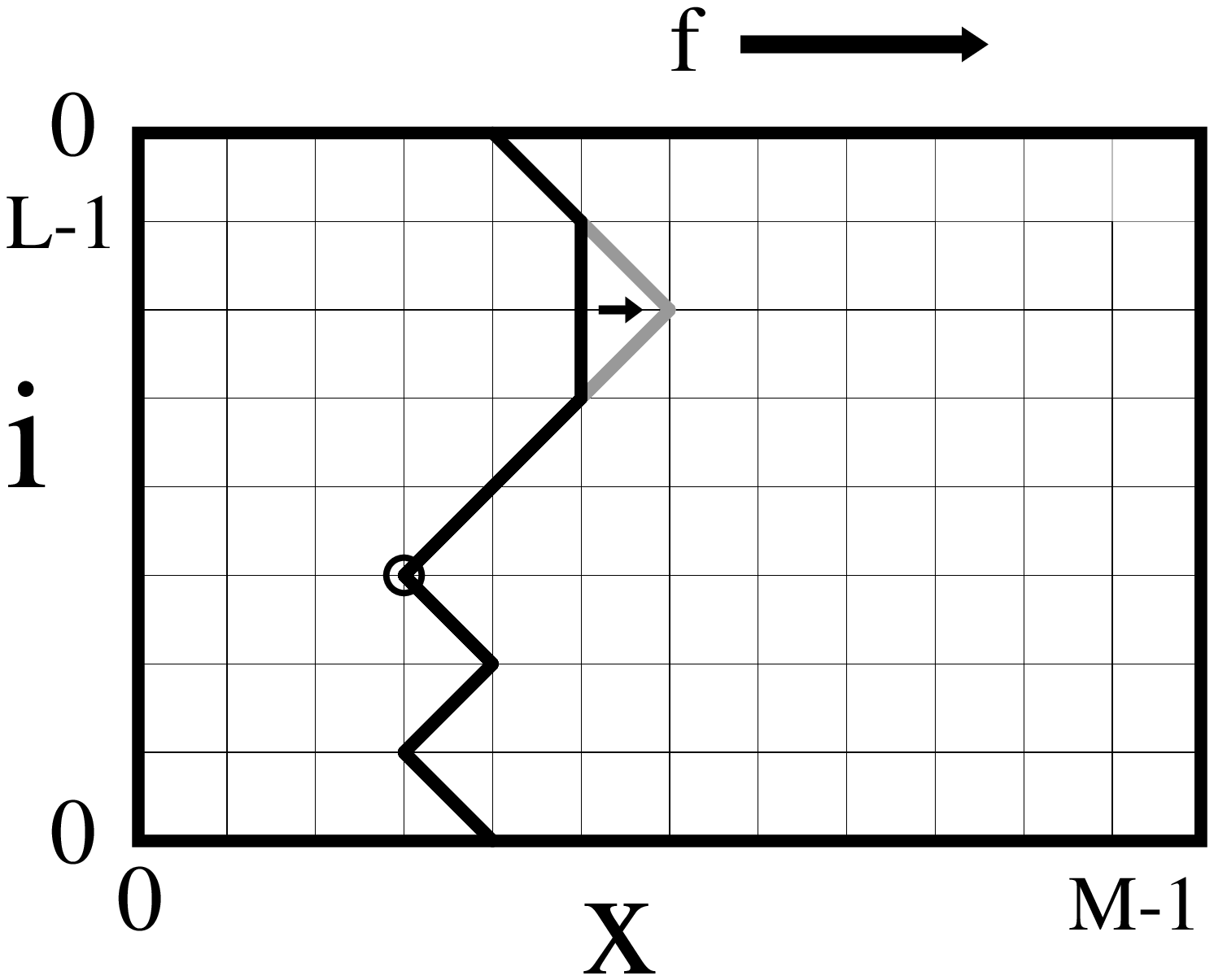,height=5.cm} }
\caption{Flux line $x^t =\{ x_i^t\}_{i=0,\ldots,L}$ on a spatial
square lattice with disorder 
potential $V(i,x)$.
Periodic boundary conditions in $i$ and $x$ are assumed
for the lattice and the disorder, respectively.  
The direction of the driving force $f$ is indicated, as well as a
proposed {\em local} Monte Carlo move. The local
dynamics lead to a trivial critical force in the limit of large
systems.}
\label{modeldef}
\end{figure}

Our model is sketched in figure \ref{modeldef}. We consider a flux
line $x^t = \{ x_i^t\}_{i=0,\ldots,L}$ moving at times $t=0,1,2,\ldots$
on a spatial square lattice with a random potential $V(i,x)$ with
$x=0,1,\ldots$.  Mainly for convenience ({\em cf} below), we also
introduce a metric constraint
\begin{equation}
 | x_{i+1}^t -x_i^t | \le 1 
\label{metric} 
\end{equation}
as well as periodic boundary conditions  ($x_0^t=x_L^t$) on the 
flux line.
The random potential satisfies
\begin{equation}
V(i,x_i+M)=V(i+L,x_i)=V(i,x_i). 
\label{periodicV} 
\end{equation} 
This condition defines an effective sample of size $(L,M)$.  While
eq.(\ref{periodicV}) is essential for the following, the periodic
boundary condition for the flux line and the specific choice of
lattice are inessential details.

The energy of a flux line $x^t$ in presence of an
external driving force $f$   is given by 
\begin{equation}
E(x^t)=\sum_{i=1}^{L}V(i,x_i^t)-f x_i^t
+\frac {c}{2}(x_{i+1}^t-x_i^t)^2 ,
\label{energy} 
\end{equation}
where $c$ is an elastic constant.

In figure \ref{modeldef}, a local Monte Carlo move is indicated.
In the local Monte Carlo algorithm, the proposed configuration
$\tilde{x}$ differs from the present configuration $x^t$ only on
a random position $i$. One chooses $\tilde{x_i} =x_i\pm 1$ with
equal probability. At zero temperature, the move is accepted
($x^{t+1}=\tilde{x}$) if the energy eq.(\ref{energy}) decreases
and if the metric constraint eq.(\ref{metric}) is satisfied.
Otherwise, it is rejected  ($x^{t+1}=x^t$).  

This rule has been used in past simulations, in spite of its very
serious shortcomings.  Consider, for example, the site in figure
\ref{modeldef} marked with a circle, $(i_0,x_0)$.  The flux line
$x^t$ shown in figure \ref{modeldef} can only move away from
$(i_0,x_0)$ if $f > V(i_0,x_0+1) - V(i_0,x_0) +c$.  Even an infinitely
long flux line $(L \rightarrow \infty)$  is thus stopped by a single
deep pin $V(i_0,x_0)$
and the motion does not differ qualitatively from the one of a
point in a disordered potential \cite{Derrida,Vinokur}.  For an
unbounded distribution of $V$, the critical force is always infinite,
exposing clearly the pathology of the local Monte Carlo algorithm.

Some authors have therefore used a bounded distribution, $V<V_{\max}$.
However, it is easy to see that in this case the local Monte Carlo
algorithm does not correctly incorporate the disorder, and the flux
line's motion is trivial: in the limit of large system size $L M
\gg 1$, the flux line will be free to move if $f> 2 V_{\max}$, in
a way which  will be similar to the motion of the non-disordered
system.  This has already been pointed out for the analogous case
of the random Ising model  \cite{Roters1}.

We conclude that the description  of a driven flux line by means
of a local Monte Carlo algorithm or its variants \cite{Yoshino}
eliminates the very  feature  which makes the problem interesting
in the first place, namely the competition between elasticity and
disorder. This competition is preserved   in the continuum Langevin
dynamics \cite{Nattermann1,Chauve}.

Within the Monte Carlo method, we are thus naturally lead to consider
generalizations of the model.  It can be seen easily that the
problem just discussed persists even if we abandon the metric
constraint eq.(\ref{metric}).  The only remaining option is therefore
to abandon the local moves  in favor of rules which allow global
moves.  The study of global moves in dynamical Monte Carlo is the
subject of this paper.

\begin{figure}
\centerline{ \psfig{figure=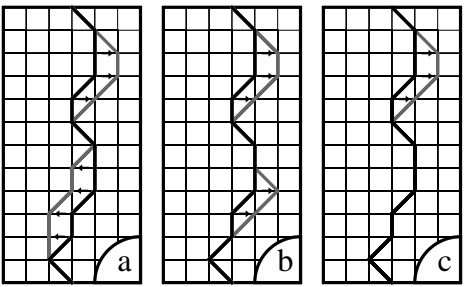,height=4.0cm} }
\caption{Non-local Monte Carlo moves which are considered in this
paper.  Model $a$: {\em all} positions $x_i^t$ ($i=1,L$) may change
at the same time by  a value $\pm 1$.  Model $b$: as in $a$, but
the motion is either in forward or in backward direction.  Model
$c$: as in $b$, but the motion is restricted  to  single `fronts'.
 }
\label{movesketch}
\end{figure}
Let us define ``Model $a$'' dynamics by a proposed move $x^t \rightarrow
x^t + \delta^t$ 
with  $\delta^t  = \{ \delta_i^t\}_{i=0,\ldots,L}$ such that
\begin{equation}
   \delta_i  = \left\{ \begin{array}{ccc}
                      +1 &                    & p\\    
                      0  &\mbox{ with prob.}  & 1 - 2 p\\    
                      -1 &                    & p
                          \end{array}
                  \right\}   
\;\;\;\forall i\;\;\;\mbox{(Model $a$)} 
\label{Modela} 
\end{equation}
At zero temperature, the proposed move is accepted, $x^{t+1} =
x^{t} + \delta $, if the resulting configuration both satisfies
the metric constraint eq.(\ref{metric}) and decreases the string
energy eq.(\ref{energy}). Note that under Model $a$ dynamics a move
is proposed with the same probability as its inverse. This serves
to enforce detailed balance, which allows to naturally extend the
rule to finite temperatures via the Metropolis algorithm. The same
can  usually not be done for cellular automata methods
\cite{Nattermann1,Nattermann2,Leschhorn}.

A possible  second rule (``Model $b$'') chooses  at each time $t$ 
with equal probability 
either to move forward  ($\mu^t =1$) or backwards ($\mu^t = -1$).  
The following move is then proposed:
\begin{equation}
   \delta_i  = \left\{ \begin{array}{ccc}
                      \mu^t&                    & p     \\
                         &\mbox{ with prob.}  &       \\
                      0  &                    & 1 -  p\\
                          \end{array}
                  \right\}
\;\;\;\forall i\;\;\;\mbox{(Model $b$)} 
\label{Modelb} 
\end{equation}

The simulation of dynamical models with such global moves may appear
hopeless because of the difficulty to detect the few energetically
favorable choices among the exponential  number of possibilities
in eq.(\ref{Modela}) or in eq.(\ref{Modelb}).

To show that the situation is much less desperate, let us first
define a `forward front' of length $k$ as a contiguous set of points
$i,i+1,....,i+k-1$ which may advance together without violating
the metric constraints eq.(\ref{metric})
($\delta_{i}=\delta_{i+1}\ldots=\delta_{i+k-1} =1$ with $\delta_{i-1}
\neq 1$, $\delta_{i+k} \neq 1$). A `backward front' is defined
analogously. We call `unstable' a front which lowers the energy
eq.(\ref{energy}).  The moves proposed in figure \ref{movesketch}$a$
and \ref{movesketch}$b$ each consist of {\em two} fronts. At least
one of these must be unstable if the move is to be accepted (this
is immediately apparent for Model $b$ and follows for Model $a$
from an elementary consideration).  To determine whether a
configuration $x^t$ is unstable, we only need to consider the at
most  $2L(L-1)+2$ fronts of $x^t$ rather than the exponential number
of moves in eq.(\ref{Modela}) or eq.(\ref{Modelb}).

Besides Model $a$ and Model $b$ dynamics, it is also possible to
set up single-front dynamical rules which respect detailed balance.
These rules (as sketched in figure \ref{movesketch} $c$)  can be
simulated with less effort than Model $a$ or Model $b$. Even in
the latter cases, though, we have developed methods which realize
eqs (\ref{Modela}) and (\ref{Modelb}) while never attempting  a
move forbidden by eq.(\ref{metric}). The question of which dynamical
rule is most satisfactory both from physical and computational
viewpoints will be dealt with elsewhere \cite{RossoKrauth2}.

Our main point in the present paper is that a great deal of
information is available without  actually simulating the dynamic
rules. We will show that  $f_c$ is the same for all models and that
the critical flux lines can be obtained easily.

We define, for an arbitrary flux line $x^{\alpha}$, the `depinning
force' $f_d(x^{\alpha})$ as the smallest non-negative  $f$ in
eq.(\ref{energy}) which destabilizes one forward front.  Furthermore,
we define the critical force of a given sample (of size $L \times
M$) as \begin{equation} f_c = \max_{ \{  x^{\alpha} \} } f_d(x^{\alpha})
\label{fcrit} \end{equation} where $\{  x^{\alpha} \}$ is the set
of all possible flux lines.  Notice  that the definiton of $f_c$
or $f_d(x^{\alpha})$ is model-independent.  We show in the following
that $f_c$ in  eq.(\ref{fcrit}) is an appropriate definition for
all cases as, for a driving force $f < f_c$, the system will be
pinned in the long-time limit $t \rightarrow \infty$.

To prove the above, we introduce a Variant Monte Carlo (VMC)
algorithm which, as a byproduct, will allow us to actually compute
$f_c$ with great ease.

At each time-step $t=0,1,\ldots$, the VMC algorithm simply moves a
single front of {\em minimal} length $k$ among the unstable forward
and backward fronts. The VMC method violates detailed balance and
is therefore not a valid Monte Carlo algorithm.  However, each move
possible within the VMC algorithm is also allowed with all the
other models considered.

We have proven the following theorem: if, under VMC dynamics at
driving force $f$, a flux line $x^{\alpha}$ is pinned in forward
direction, it can at most  recede towards a configuration $x^{\beta}$
($x_i^{\beta} \le x_i^{\alpha}  \; \forall i$), which is itself
pinned in forward direction. Eventually, we will reach a flux line
$x^{\gamma}$ which is pinned both in forward and in backward
directions.  This flux line $x^{\gamma}$ is pinned for all models;
if it is pinned at $f_c$, we call it a `critical flux line' $x^{c}$.

The theorem allows us to understand  that eq.(\ref{fcrit}) is indeed
an  appropriate definition for all models: As we defined
$f_d(x^{\alpha})$ only with respect to {\em forward} motion, one
might have imagined that a flux line which cannot advance at $f_c$
could move backwards and then  be avoided during the subsequent
forward evolution.  Our theorem tells us that such loopholes do
not exist: Under VMC dynamics, a flux line which can no longer move
forward, will move backwards and then stop.

Conversely, we can show that a flux line which can no longer move
backwards under the VMC dynamics will exclusively move forward and
then stop.  This observation simplifies the numerical computations
of the VMC algorithm.

Now, we treat the question of how to determine $f_c$ and $x^c$.
There is no a priori guarantee that a generic dynamic rule (such
as  Model $a$ or Model $b$) will actually encounter an $x^{c}$,
when driven at forces $f \leq f_c$.  Simulations in small systems,
where $f_c$ and  $x_c$ can be obtained by exact enumeration, show
that $x^t$ may pass the sample many millions of times without
getting pinned.  We initially even suspected that $x^{c}$ could be
dynamically inaccessible from part of phase space.

In this context, we have proven  a second theorem: Starting at an
initial configuration $x^{t_0}$ with $ x^{t_0}_i \le x^c_i$, the
VMC algorithm at driving force $f \le f_c$  can never pass $x^c$.
In practice, we simply update the driving force by the present
depinning force $f=f_d(x^t)$ each time we get stuck at a configuration
$x^t$. In one pass through the system, we will have obtained the
critical force.  The computation of $f_c$ and of the critical flux
line is thus extremely simple.

Furthermore, the
VMC algorithm gives an explicit construction - for any of the
methods - which dynamically connects an arbitrary initial state
with a critical flux line. This proves that all the models in figure
\ref{movesketch} possess a single ergodic component.

We now present our numerical calculations which show that $f_c$
is finite for a sample of size $(L,M)$ in the limit $L,M \rightarrow \infty$
with $L/M= const$.  In all our calculations, we have used a Gaussian
normal distribution for the random potential.

For finite sizes $(L,M)$, we define  the integrated distribution
function $P_{L,M}(f)$ as the probability that a sample
of size $(L,M)$ possesses a critical force $f_c \le f$. Because of
the metric constraint eq.(\ref{metric}), we know that
the lateral extension of the flux line,  
$\max_{i,j}| x_i - x_j|$,   will be at most $L/2$. This can be used
to show that for $M \gg L/2 $ 
\begin{equation}
P_{L,2 M}(f) =  P_{L,M}^2(f).   
\label{scaling} 
\end{equation}
\begin{figure}
\centerline{ \psfig{figure=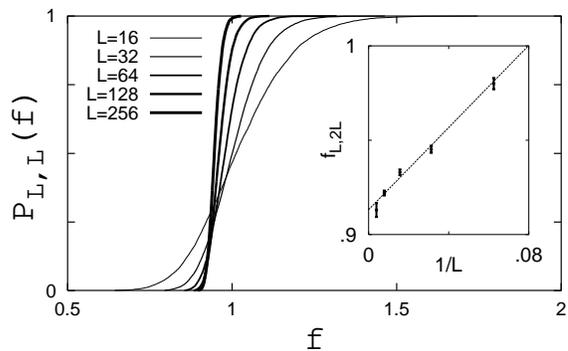,height=5 cm} }
\caption{Main figure: Probability $P_{L,L}(f)$ to have a critical
force $f_c < f$ {\em vs} $f$ for samples of size $(L,L)$ {\em vs} $L$
for $L=16,32,\ldots,512$. The elastic constant is $c/2=1$, and a
Gaussian normal distribution is used for the disorder potential.
For all large $M$, the curves $P_{L,M}(f)$ and $P_{2L,2M}(f)$
intersect at the same force, $f_{L,2L}$ which,  in the inset, is plotted
{\em vs} $1/L$. The extrapolated value of $f_{L,2L}$ in the limit
$1/L  \rightarrow 0$ is the critical force $f^{\infty}_c$ of a 
macroscopic sample.}
\label{datacollaps}
\end{figure}

We will be interested in the intersection point, $f_{L,2L}$, between 
the integrated probability distribution for the system of size $(L,M)$ 
and the one of size $(2L,2M)$: 
\begin{equation}
P_{L,M}(f_{L,2L}) = P_{2 L,2 M}(f_{L,2L}).
\end{equation}
In fact, 
$f_{L,2L}$ will not depend on $M$  for large
$M$, by virtue of eq.(\ref{scaling}). In our opinion, this observation
implies that the natural scaling  for our system in the thermodynamic
limit is $M \sim L^{\gamma}$   with $\gamma = 1$, {\em i.e.} that we
should compare the system of size $(L,M)$ with another one, double in
size both in $L$ and in $M$.

We have checked numerically that corrections to the scaling relation
eq.(\ref{scaling})  are already negligible for $L\sim M$ (for $L
> 4$) and that intersection points $f_{L,2L}$ indeed do not depend
on $M > L$.  In figure \ref{datacollaps} we show data
for $P_{L,L}(f)$ for $L=16,32,\ldots,512$.
The inset of the figure gives the  $f_{L,2L}$ as a function of
$1/L$ for all sizes. It is evident that 
$f_{L,2L}$ extrapolates to a finite value, the critical force of
the model in the thermodynamic limit, $f_c^{\infty}$. We find $f_{L,2L} 
\rightarrow f^{\infty}_c = 0.91 \pm 0.01$ for
$c/2=1$. We stress again that $f^{\infty}_c$ is independent of the aspect
ratio $L/M$.
  
For the critical flux line $x_c$, we also studied the size $k$ of
the minimal unstable front. Naturally, the probability distribution
of $k$, $p(k)$, is unbounded, as a finite support of $p(k)$ would
lead us back to the inconsistencies of the local Monte Carlo
algorithm.  For large $k$, we find an exponential distribution
$p(k) \sim \exp(-k/k_{typ})$ where $k_{typ}$ depends on the elasticity
parameter $c$, but remains finite as $L\rightarrow \infty$. Of
course, this initial seed of the motion beyond $x_c$ may well
trigger motions on much larger scales. These problems will be
studied separately \cite{RossoKrauth2}.

Finally, we discuss possible extensions of the work presented here.
We  already indicated that our metric constraint  was introduced mainly
for convenience and that  all the developments remain valid. In the
absence of the constraint, the lateral extension of the flux line
may however  scale as $L^{\gamma}$ with $\gamma > 1$. If so, our
scaling assumption eq.(\ref{scaling}) would have to be modified.
We have also extended most of our results to higher-dimensional
manifolds and embedding spaces. There, the only critical issue
seems to be the  complexity of the VMC algorithm, as the number of
possible fronts can be much larger than in the linear flux line.

In conclusion, we have put the dynamical Monte Carlo algorithm for
the motion of elastic manifolds in random media on a solid footing.
We have shown that only global-move schemes can capture the subtle
interplay between elasticity and disorder, which is totally absent
from the customary local algorithms. Our theorems allowed us to
compute features universal to all members of this class, namely
the critical force, as well as properties of the critical flux
lines.  The variant Monte Carlo algorithm is crucial in that it
allows to compute the critical force with full rigor even for
samples which are several orders of magnitude larger than those
accessible to exact enumeration methods.

In the future, we think that the extensions to higher dimensions
will be a very interesting subject for further study. Another
challenge will be  to understand the actual time dependence, both
at zero and at finite temperatures.  It would also be very important
to have rigorous mathematical proofs concerning the finiteness of
the critical force and the  scaling, in various dimensions. We are
convinced that the VMC algorithm is sufficiently simple to allow
such an approach.

Acknowledgments: It is a pleasure to thank P. Chauve, P. Le Doussal and
L. Santen for very helpful discussions.


\end{multicols}

\end{document}